\documentclass[
journal=jpccck,
layout=twocolumn
]{achemso}

\SectionNumbersOn
\usepackage[version=3]{mhchem} 
\usepackage{mathcomp} 
\usepackage{braket}
\usepackage{CJKutf8} 
\usepackage{booktabs}

\makeatletter
\renewcommand\@seccntformat[1]{\csname the#1\endcsname.\quad}
\makeatother

\usepackage{xcolor}
\usepackage[normalem]{ulem}

\newcommand{\Add}[1]{\textcolor{black}{#1}}	
\newcommand{\Erase}[1]{\if0{#1}\fi}	

\newcommand{\AddII}[1]{\textcolor{black}{#1}}	
\newcommand{\EraseII}[1]{\if0{#1}\fi}	
\newcommand{\ReviseII}[2]{\textcolor{black}{#2}} 

\author{Ryotaro Nakazawa}
\email{nakazawa@ims.ac.jp}
\phone{+81 90-564-7445}
\affiliation[IMS]
{Institute for Molecular Science, Okazaki 444-8585, Japan}
\affiliation[Chiba Uni GSSE]
{Graduate School of Science and Engineering, Chiba University, Chiba 263-8522, Japan}

\author{Masaya Kitaoka}
\affiliation[Chiba Uni GSSE]
{Graduate School of Science and Engineering, Chiba University, Chiba 263-8522, Japan}

\author{Ryota Kaimori}
\affiliation[Chiba Uni GSSE]
{Graduate School of Science and Engineering, Chiba University, Chiba 263-8522, Japan}

\author{Manato Tateno}
\affiliation[Chiba Uni GSSE]
{Graduate School of Science and Engineering, Chiba University, Chiba 263-8522, Japan}

\author{Runa Hoshikawa}
\affiliation[Chiba Uni GSSE]
{Graduate School of Science and Engineering, Chiba University, Chiba 263-8522, Japan}

\author{Yuya Tanaka}
\affiliation[Gunma Uni]
{Graduate School of Science and Technology, Gunma University, Kiryu 376-8515, Japan}

\author{Hisao Ishii}
\affiliation[Chiba Uni GSSE]
{Graduate School of Science and Engineering, Chiba University, Chiba 263-8522, Japan}
\alsoaffiliation[Chiba Uni CFS]
{Center for Frontier Science, Chiba University, Chiba 263-8522, Japan}
\alsoaffiliation[Chiba Uni MCRC]
{Molecular Chirality Research Center, Chiba University, Chiba 263-8522, Japan}
 
\title[An \textsf{achemso} demo]
  {Determining the Density of In-Gap States in Organic Semiconductors: A Pitfall of Photoelectron Yield Spectroscopy}

\begin{document}
\begin{CJK*}{UTF8}{min} 

\begin{abstract} 
Accurate determination of low-density electronic states in the bandgap (in-gap states) is crucial for optimizing the performance of organic optoelectronic devices. Derivative photoelectron yield spectroscopy (PYS) is employed to estimate the density of states (DOS) of in-gap states. However, low-energy photons in PYS can generate excitons and anions in organic semiconductors, raising questions about whether derivative PYS spectra truly represent the DOS. 
We revealed that PYS signals originate from the single-quantum external photoelectron effect (SQEPE) of in-gap states, SQEPE of the singly occupied molecular orbital (SOMO) of anions, and the biphotonic electron emission (BEE) effect via exciton fusion. 
Because BEE signals mask the DOS contribution, derivative PYS misestimates the DOS of in-gap states. In contrast, constant final state yield spectroscopy (CFS-YS) reliably determines the DOS by separating these components. For a tris(8-hydroxyquinoline) aluminum (Alq$_3$) film, CFS-YS revealed the DOS of in-gap and SOMO states over six orders of magnitude, clarifying why the Alq$_3$ layer works effectively in organic light-emitting diodes. In the devices, BEE can act as \Erase{carrier-generation and }degradation processes, and CFS-YS \Add{is a useful method to} also probe it. 
We provide the practical guidelines of low-energy photon measurements for DOS determination, such as measurements of photon-flux dependency. 
\end{abstract}

\section{Introduction}
Organic semiconductors are widely used in optoelectronic devices such as organic solar cells and organic light-emitting diodes (OLEDs) owing to their flexibility, environmental compatibility, low weight, and material diversity\cite{Chen2023_OPV_intro}.
The operating principles of these devices are typically described with energy diagrams derived from the materials’ ionization energy and electron affinity. However, such simplified diagrams neglect a small but significant density of states (DOS) extending into the bandgap from the highest occupied molecular orbital (HOMO) and lowest unoccupied molecular orbital (LUMO).
These electronic states within the bandgap are commonly termed in-gap or tail states.
Simulations indicate that this DOS tailing substantially contributes to the vacuum-level shift at metal/organic interfaces\cite{Oehzelt2014_VLshift_simulation}.
Charge injection from electrodes into the organic layer is reported to arise primarily from the DOS of in-gap states rather than from the HOMO\cite{Shimizu2022_electroninjection}. Experimental studies have further shown that these in-gap states induce vacuum-level shifts at metal/organic interfaces\cite{Yonezawa2014_in-gap_state_HSUPS, Hagenlocher2021_in-gap_state_HSUPS}.
Accurate observation and evaluation of the DOS of in-gap states are therefore essential for optimizing the electronic characteristics of semiconductor devices.

Photoelectron yield spectroscopy (PYS) is widely employed to evaluate occupied electronic states in device-related materials because of its measurement simplicity. In this method, the sample is irradiated with monochromatic photons from ultraviolet to visible wavelengths while varying the photon energy, and the total photoelectron yield is typically recorded with a channeltron under high vacuum.
For ambient-pressure measurements, several techniques have been developed in which the total yield is detected either as ionized O$_2$ molecules using an open-counter analyzer\cite{Kirihata1981_PYS_opencounter} or as an electric current (photo-holes) using an ammeter\cite{Inumaru1999_PYS_current, Nakayama2008_pys_current}.
PYS is commonly applied to determine the ionization energy of semiconductors and insulators\cite{pope1999electronic_BEE_book, Nakayama2008_pys_current, Ishii2015_PYSreview_book_section, Murata1979_PYS}.
Moreover, derivative PYS (DPYS) spectra have been used to estimate the DOS distribution of occupied in-gap states in inorganic and organic semiconductors\cite{Sebenne1977_PYS_differentiation, Nakano2021_PYS_differentiation, Miyazaki1999_PYS_gap_states}, assuming that the PYS spectrum approximates the integral of the initial DOS (see \ReviseII{Theory section}{Section \ref{sec:theory}}).
Tadano \textit{et al.} suggested that applying the second derivative of the PYS spectrum enhances the accuracy of DOS evaluation\cite{Tadano2019_PY2DS}, and this method has also been used for DOS determination\cite{Eguchi2023_secondderivativePYS, Eguchi2021_secondderivativePYS}.

However, the interpretation of PYS spectra in organic semiconductors remains uncertain.
Because PYS employs low-energy photons from ultraviolet to visible wavelengths, the measurement photons can effectively generate exciton and anion states.
Kinjo \textit{et al.} reported a photoelectron signal arising from anion states in thin films of polar molecules\cite{Kinjo2016_negativeion}, while Pope \textit{et al.} and Kotani \textit{et al.} observed biphotonic electron emission (BEE) caused by exciton--exciton and exciton--free electron fusion in several typical organic crystals\cite{pope1999electronic_BEE_book, Ono2001_BEE}. These findings indicate that PYS spectra should not be interpreted solely as single-quantum external photoelectric effects (SQEPE) of occupied in-gap states.
Accordingly, the validity of using DPYS or the second derivative of PYS as a direct measure of the DOS of occupied in-gap states requires careful re-evaluation.

Constant final state yield spectroscopy (CFS-YS) measures the partial photoelectron yield at a fixed kinetic energy while varying the photon energy from ultraviolet to visible light, enabling determination of the DOS in valence-band and occupied in-gap states\cite{Sebastiani1995CFSYS,Korte2006_CFSYS_Si, Levine2021_CFSYS_pero, Menzel2022_CFSYS_pero, Nakazawa2021_CFSYS_aIGZOasdepo, Nakazawa2024_IGZO, Shimizu2022_electroninjection, Inoue2025_CU2O_CFSYS}.
Because CFS-YS employs low-energy photons, similar to PYS, photoelectrons may arise not only from the SQEPE of occupied in-gap states but also from SQEPE of anion states and BEE effect in organic semiconductors.
Therefore, verifying whether the CFS-YS spectrum truly reflects the DOS of occupied states is essential.

To elucidate the origin of photoelectrons detected by PYS and CFS-YS, measuring their energy distribution curves is effective. Sato \textit{et al.} developed $h\nu$-dependent high-sensitivity ultraviolet photoelectron spectroscopy (HS-UPS), which acquires a series of energy distribution curves by varying photon energies from ultraviolet to visible light to observe occupied in-gap states\cite{Sato2017_HSUPS_1015cm-3eV-1}.
This approach clarifies the origin of photoelectrons observed in both CFS-YS and PYS.

In this study, we examine whether photoelectron techniques employing low-energy photons can accurately determine the DOS of in-gap states in organic semiconductors.
We show that such photons can generate photoelectrons through the SQEPE of occupied in-gap states, SQEPE of anion states, and BEE effect via exciton fusion.
If DOS- and BEE-derived signals are detected, the DOS obtained from DPYS differs from the true DOS because PYS cannot distinguish these contributions.
Accurate DOS determination therefore requires identifying the photoelectron origin with $h\nu$-dependent HS-UPS, and performing CFS-YS at an appropriate kinetic energy enables the reliable DOS determination by suppressing BEE contributions.
This work provides guidelines for reliable DOS evaluation of in-gap states via photoelectron spectroscopies by using low-energy photons.
While UPS is a specialized technique, PYS is widely adopted in device studies; our recommendations support both communities in achieving reproducible DOS measurements.
\section{Method}
\subsection{Theoretical basis}\label{sec:theory}
When photons irradiate a sample, some electrons are emitted from within the solid into the vacuum.
These emitted electrons, termed photoelectrons, are measured as a function of kinetic energy by photoelectron spectroscopy. Photoelectrons generated by the SQEPE of occupied states are described by the \textit{three-step model}\cite{pope1999electronic_BEE_book}. According to this model, the lineshape of the angle-integrated photoelectron spectrum $N_\mathrm{PES}$ is expressed as\begin{align}
N(E_\mathrm{k}, h\nu)_\mathrm{PES}
&\propto (h\nu) \lvert M_{\mathrm{fi}}\rvert^{2} 
D_\mathrm{i}(E_\mathrm{k} - h\nu) \nonumber \\
&\quad \ \ D_\mathrm{f}(E_\mathrm{k}) \,
X(E_\mathrm{k}) \,
T(E_\mathrm{k}).
\label{eq:HSUPS}
\end{align}
Here, $h\nu$ and $E_\mathrm{k}$ denote the incident photon energy and the kinetic energy of the photoelectrons, respectively. The quantity $(E_\mathrm{k}-h\nu)$ corresponds to the binding energy with the vacuum level set to 0 eV. 
$\lvert \textit{M}_{\mathrm{fi}}\rvert^{2} = \lvert 
\Braket{\psi_\mathrm{f}|\boldsymbol{r}|\psi_\mathrm{i}} \rvert^{2}$ represents the squre of matrix elements from an initial to a final state, where $\psi_\mathrm{f}$ and $\psi_\mathrm{i}$ are the final and initial wave functions, respectively, and $\boldsymbol{r}$ is the position vector. 
$D_\mathrm{i}$ and $D_\mathrm{f}$ denote the initial and final DOS, while $X$ and $T$ represent the transport probability to the surface and the transmittance probability through the surface, respectively.
In PYS measurements, the total photoelectron yield $Y_\mathrm{PYS}$ is recorded as a function of the incident photon energy $h\nu$.
When photoelectrons originate from the SQEPE of occupied states, $Y_\mathrm{PYS}$ is expressed according to Eq. \ref{eq:HSUPS} as
\begin{equation}
\begin{aligned}
Y_\mathrm{PYS}(h\nu) 
&\propto 
\int_{0}^{h\nu} N_\mathrm{PES}(E_\mathrm{k}, h\nu) \, dE_\mathrm{k} \\
&= \int_{0}^{h\nu} \!(h\nu)
\lvert M_{\mathrm{fi}}\rvert^{2} \,
D_\mathrm{i}(E_\mathrm{k}-h\nu) \\
&\quad \quad \quad D_\mathrm{f}(E_\mathrm{k}) \,
X(E_\mathrm{k}) \,
T(E_\mathrm{k}) \, dE_\mathrm{k}.
\end{aligned}
\label{eq:PYS}
\end{equation}
Here, if
$(h\nu)\lvert \textit{M}_{\mathrm{fi}}\rvert^{2}$,
$D_\mathrm{f}(E_\mathrm{k})$,
$X(E_\mathrm{k})$,
and $T(E_\mathrm{k})$
are assumed constant with respect to photon energy, Eq. \ref{eq:PYS} simplifies to 
\begin{equation}
\begin{aligned}
 Y_\mathrm{PYS}(h\nu) 
 \propto 
 \int_0^{h\nu} 
 D_\mathrm{i}(E_\mathrm{k}-h\nu)
 dE_\mathrm{k}.
  \label{eq:PYS_int_DidE}
\end{aligned}
\end{equation}
Under this assumption, the derivative of the PYS (DPYS) spectrum, $dY_\mathrm{PYS}/d(h\nu)$, corresponds to $D_\mathrm{i}(E_\mathrm{k}-h\nu)$\cite{Tadano2019_PY2DS, Sebenne1977_PYS_differentiation, Nakano2021_PYS_differentiation}.
However, when in-gap states are probed, low-energy photoelectrons at approximately 1--2 eV above the vacuum level are analyzed. In this regime, $T(E_\mathrm{k})$ can be approximated to increase monotonically with $\textit{E}_\mathrm{k}$, based on the semiclassical \textit{escape-cone argument}\cite{Hufner2003_photoemission_spectroscopy}. Tadano \textit{et al.}\cite{Tadano2019_PY2DS} refined this approximation by assuming $T(E_\mathrm{k}) = E$.
Under these conditions, $N_\mathrm{PYS}$ is expressed as\[
 Y_\mathrm{PYS}(h\nu) 
 \propto 
 \int_0^{h\nu} 
 D_\mathrm{i}(E_\mathrm{k}-h\nu)
 E_\mathrm{k}
 dE_\mathrm{k}.
\]
Hence, $D_\mathrm{i}(E_\mathrm{k}-h\nu)$ can be approximated by the second derivative of the PYS spectrum, $d^2Y_\mathrm{PYS}/d(h\nu)^2$, as detailed in the literature\cite{Tadano2019_PY2DS}.

In CFS-YS measurements, the partial photoelectron yield $Y_\mathrm{CFS}$ at a fixed kinetic energy is recorded as a function of $h\nu$. In principle, $D_\mathrm{f}$, $X$, and $T$ in Eq. \ref{eq:HSUPS} are constant, allowing $Y_\mathrm{CFS}$ to be expressed as follows:
\begin{equation}
\begin{aligned}
 Y_\mathrm{CFS}(E_\mathrm{k}, h\nu) \propto (h\nu) \lvert \textit{M}_{\mathrm{fi}}\rvert^{2} D_\mathrm{i}(E_\mathrm{k}-h\nu).
\end{aligned}
\label{eq:CFSYS}
\end{equation}
$Y_\mathrm{CFS}/h\nu$ can be regarded as $D_\mathrm{i}$, when $\lvert \textit{M}_{\mathrm{fi}}\rvert^{2}$ dependence of $h\nu$ is small. 
In this article, $Y_\mathrm{CFS}/h\nu$ is referred to as the CFS-YS spectrum.
In principle, CFS-YS can determine \Add{$D_\mathrm{i}$} with fewer assumptions than the DPYS \Add{and} second derivative of PYS.

\subsection{Experimental}

Tris(8-hydroxyquinoline) aluminum (Alq$_\mathrm{3}$) and fullerene (C$_\mathrm{60}$) were deposited on indium tin oxide (ITO)-coated glass substrates via vacuum vapor deposition. The thicknesses of Alq$_3$ and C$_\mathrm{60}$ films were approximately 30 nm and 50 nm, respectively, with deposition rates of 1 and 0.4 \AA/sec. The deposition was conducted at a working pressure of $10^{-5}$ Pa.

$h\nu$-dependent HS-UPS measurements were performed at Chiba University, with details of the equipment provided elsewhere\cite{Ishii2015_PYSreview_book_section}.
Samples were transferred to the measurement chamber without air exposure after deposition.
The base and working pressures during measurement were approximately 10$^{-7}$ Pa. A D$_\mathrm{2}$ lamp (Hamamatsu Photonics K.K., L1836, 150 W) and a Xe lamp (Ushio Inc., UXL-500D, 500 W) served as light sources. The incident light was monochromatic with minimal stray light (10$^{-9}$ at 632 nm), achieved by a zero-dispersion double monochromator (Bunkoukeiki Co., Ltd., BIP-M25-GTM) covering $h\nu = 1.5$--8.4 eV.
A photomultiplier tube (Hamamatsu Photonics K.K., R6836) monitored the incident photon flux, with the excitation light incident at 55$\tcdegree$ from the surface normal.
An electrostatic hemispherical analyzer (PSP Vacuum Technology Ltd., RESOLVE 120) measured the kinetic energy distribution of photoelectrons at normal emission.
The spectrum of the Fermi edge of a gold film was used to determine the total instrumental function and Fermi level.
The total energy resolution was approximately 0.17 eV under $h\nu = 7.7$ eV irradiation at room temperature.
A bias of $-10$ V was applied to ensure that the sample vacuum level exceeded that of the analyzer.
The work function was determined from the secondary-electron cutoff in the $h\nu$-dependent HS-UPS spectrum.
CFS-YS measurements were conducted using the same setup as the $h\nu$-dependent HS-UPS, while PYS measurements employed a channeltron detector to record photoelectrons.

\section{Results \& discussion}
\subsection{Photoelectron yield spectroscopy}
\begin{figure}
\includegraphics[width=85mm]{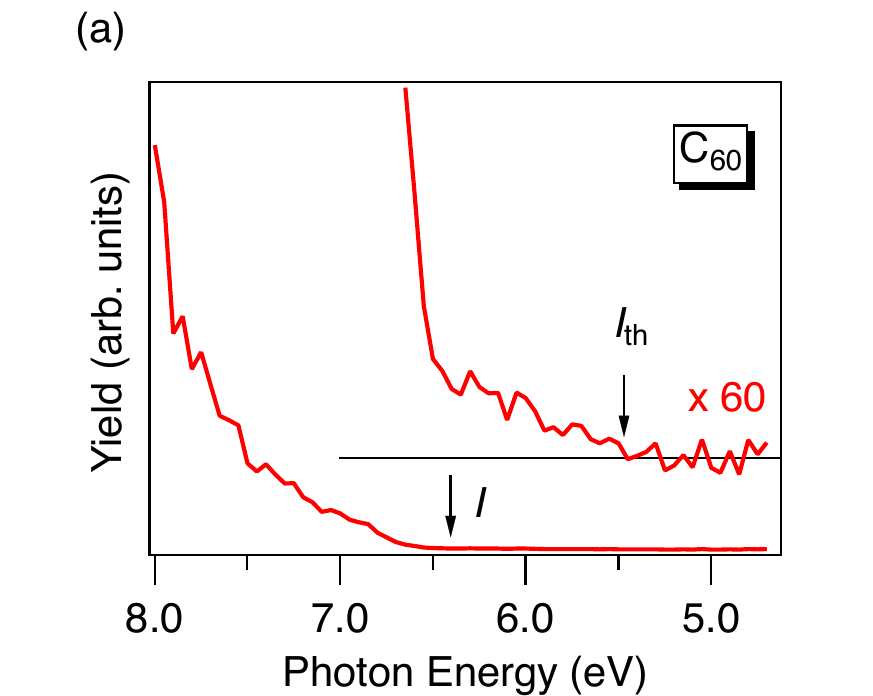}
\includegraphics[width=85mm]{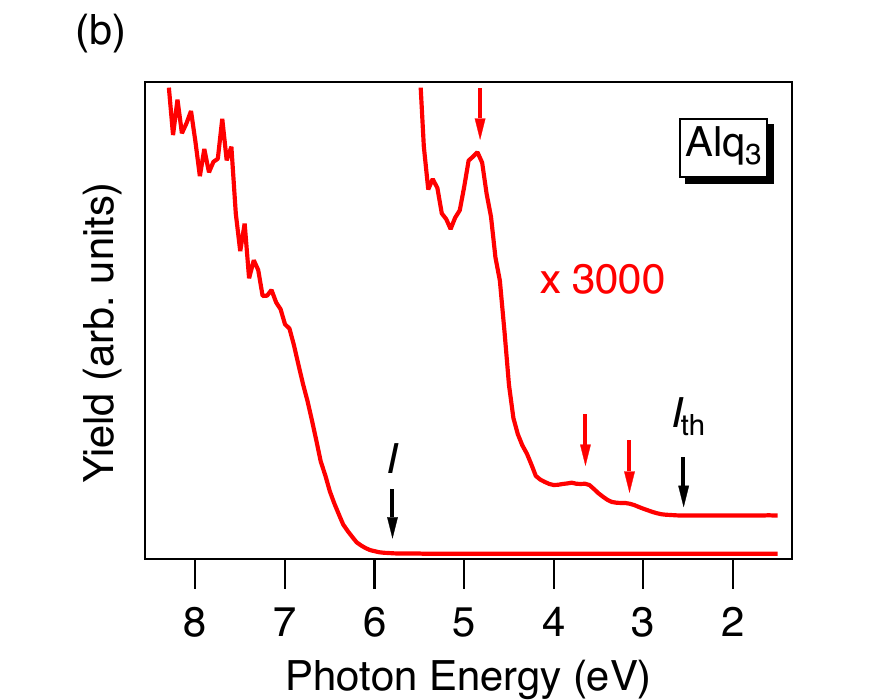}
 \caption{Photoelectron yield spectra of (a) C$_{60}$ and (b) Alq$_\mathrm{3}$ thin films. \textit{I} and $\textit{I}_\mathrm{th}$ denote the ionization energy and the subthreshold of the photoelectron yield, respectively.}
  \label{fig:PYS_Alq$_3$_Liq}
\end{figure}

Figure \ref{fig:PYS_Alq$_3$_Liq} (a) shows the PYS spectrum of the C$_\mathrm{60}$ thin film.
The spectral intensity rises for $h\nu > 6.5$ eV.
The ionization energy \textit{I} was determined by linear extrapolation of the cubic-root plot of the PYS spectrum\cite{pope1999electronic_BEE_book, Ishii2015_PYSreview_book_section}, yielding 6.3 eV, in agreement with reported values (6.36--6.5 eV)\cite{Hinderhofer2013_C60, Yamamoto2015_C60_Is, Kang2005_C60_Is}.
In the bandgap region, the subthreshold of the photoelectron yield is approximately 5.5 eV. 

Figure \ref{fig:PYS_Alq$_3$_Liq}(b) shows the PYS spectrum of the Alq$_3$ thin film.
The ionization energy \textit{I} determined by linear extrapolation of the cubic root of the PYS signal (not shown), is 5.8 eV, consistent with reported values (5.7--5.9 eV)\cite{Hill1999_alq3_Is, Hill2000_alq3_Is, Noguchi2013_interface_alq3_cath}.
Notably, several peaks appear at $h\nu = 4.9$, 3.7, and 3.2 eV (red arrows).
If the PYS signal strictly represented the integrated DOS of occupied states \Add{(Eq. \ref{eq:PYS})}, the photoelectron yield would increase monotonically with photon energy. These peaks therefore indicate that the PYS spectrum does not directly reflect the integrated occupied DOS \Add{in the low-$h\nu$ region}.\Erase{ Furthermore, although the effective work function (the difference between the vacuum level at the Alq$_3$ surface and the Fermi level of the substrate) is determined to be 3.7 eV from the secondary-electron cutoff of $h\nu$-dependent HS-UPS spectra (not shown), photoelectron yield is observed at $h\nu < 3.7$ eV. Because SQEPE from occupied states requires photon energies at least equal to the effective work function, the PYS signal below 3.7 eV cannot originate from the occupied DOS. These observations indicate that photoelectron mechanisms other than SQEPE from occupied DOS predominantly contribute to the PYS signal in the low-$h\nu$ region. 
These observation indicates that photoelectron emission mechanisms other than SQEPE from occupied DOS predominantly contribute to the PYS signal in the low-$h\nu$ region.}

\subsection{Ultraviolet photoelectron spectroscopy}\label{sec:ups}
\begin{figure}
    \includegraphics[width=80mm]{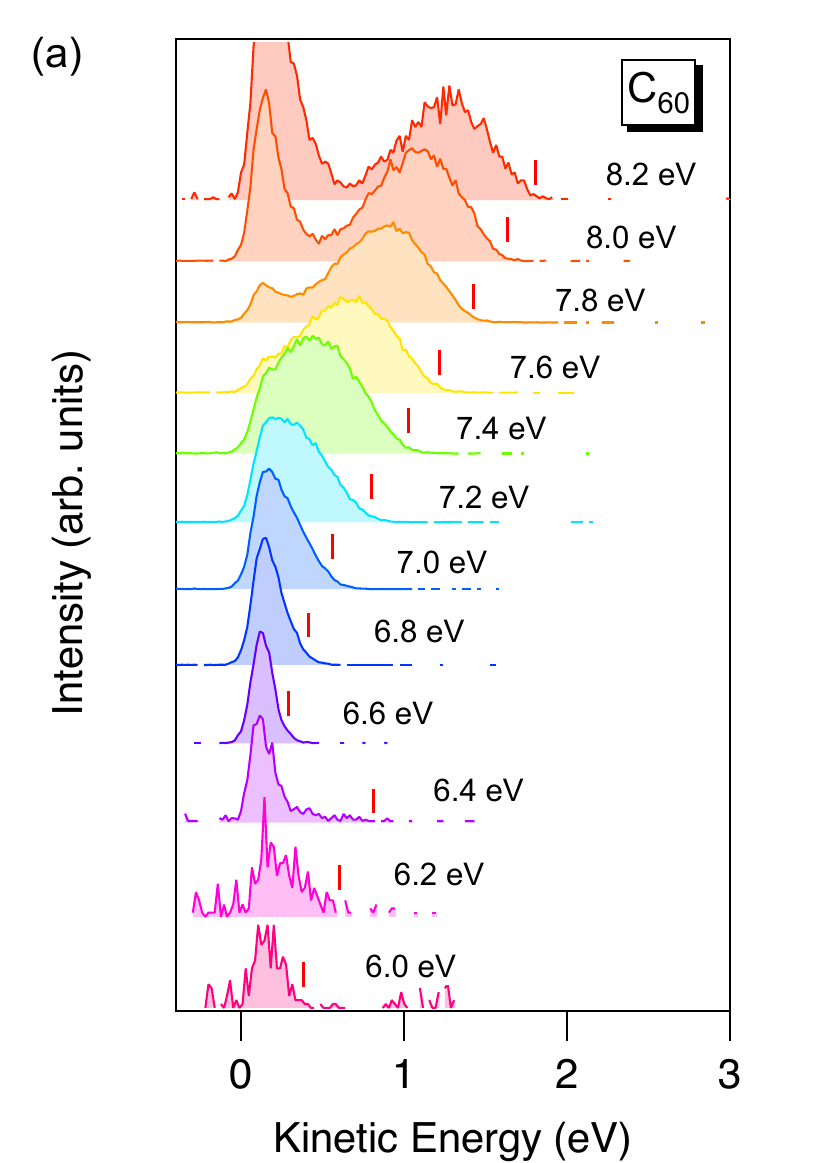}
    \includegraphics[width=80mm]{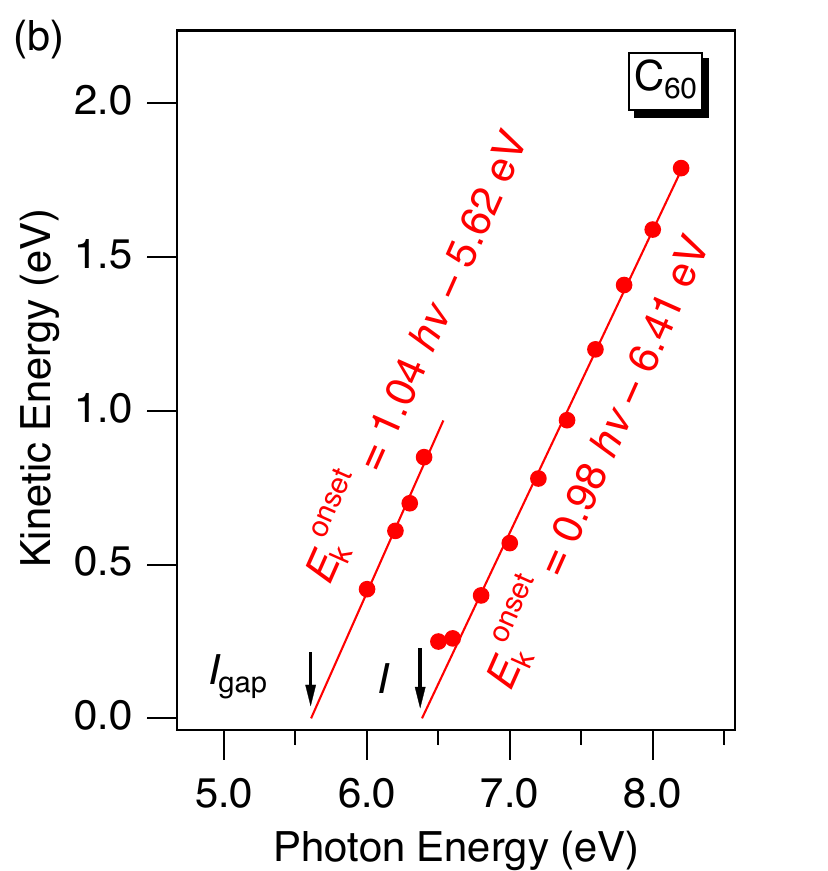}
\caption{
(a) $h\nu$-dependent HS-UPS spectra of a C$_\mathrm{60}$ thin film for $h\nu = 8.2$--6.0 eV. Intensities are normalized to the peak value. (b) Plots of the onset kinetic energy versus photon energy from the $h\nu$-dependent HS-UPS spectra. 
}
\label{fig:LEHSUPS_C60}
\end{figure}

To elucidate the origin of the PYS photoelectrons, $h\nu$-dependent HS-UPS was performed.
Figure \ref{fig:LEHSUPS_C60}(a) shows the spectra of a C$_\mathrm{60}$ thin film, with kinetic energy on the horizontal axis and the vacuum level set to 0 eV.
Spectral intensities were normalized to the maximum value in the full spectrum or the HOMO region, and the spectral onset is indicated by a bar.
As $h\nu$ decreases, the spectral onset shifts to lower kinetic energies in the ranges $h\nu = 8.2$--6.6 and 6.4--6.0 eV.
Figure \ref{fig:LEHSUPS_C60}(b) plots the onset kinetic energy ($E_\mathrm{k}^\mathrm{onset}$) versus $h\nu$, showing a linear relation with a slope near 1 for $h\nu = 8.2$--6.6 eV.
This slope is consistent with the following relation:
\begin{equation}
    E_\mathrm{k}^\mathrm{onset}=h\nu-\textit{I}.
\label{eq:SQEPEofHOMO}
\end{equation}
which describes photoelectron emission due to SQEPE from the HOMO.
Using this equation, the ionization energy is determined to be 6.27 eV. 
For $h\nu = 6$--6.4 eV, the spectral onset also exhibits a linear relation with a slope near 1, consistent with 
\begin{equation}
    E_\mathrm{k}^\mathrm{onset}=h\nu-\textit{I}_\mathrm{gap},
\label{eq:SQEPEofIngap}
\end{equation}
representing photoelectron emission from the SQEPE of occupied in-gap states.
Here, $\textit{I}_\mathrm{gap}$ denotes the ionization energy of the occupied in-gap states, determined to be 5.62 eV.

\begin{figure*}
    \centering
    \includegraphics[width=50mm]{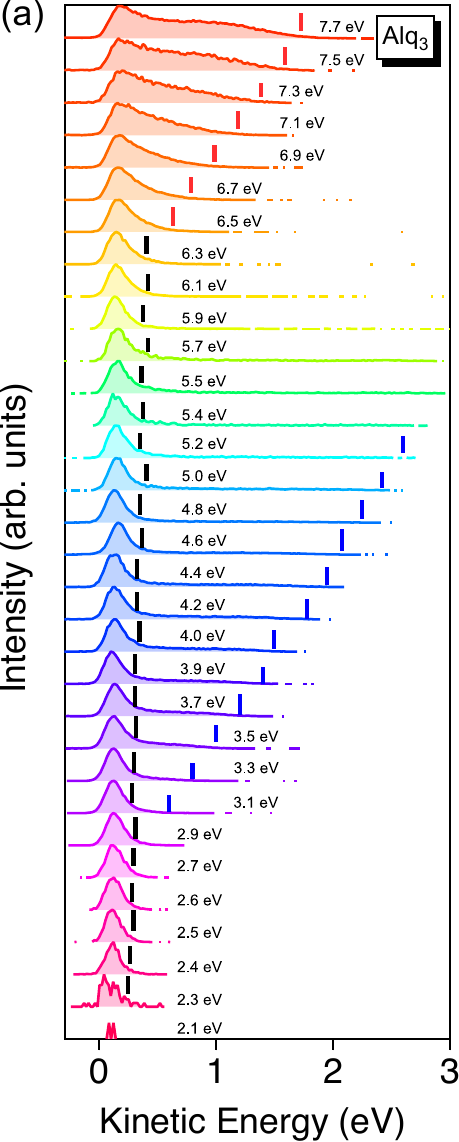}
    \includegraphics[width=45mm]{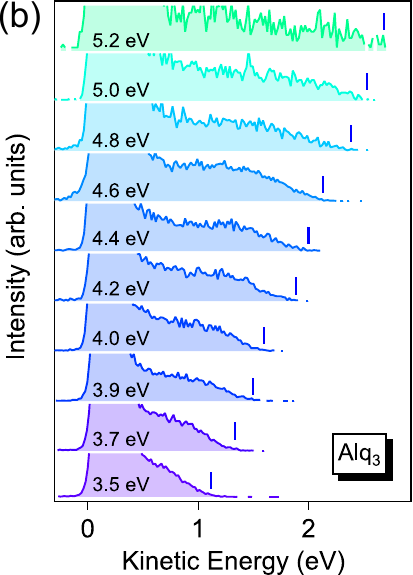}
    \includegraphics[width=50mm]{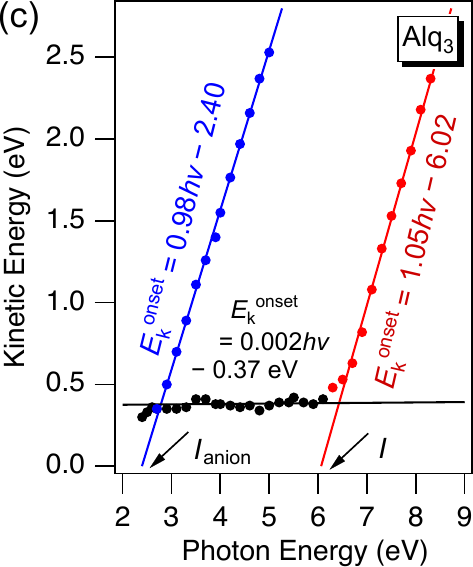}
\caption{
(a) $h\nu$-dependent HS-UPS spectra of the Alq$_3$ thin film for $h\nu = 7.7$--2.3 eV. Intensities are normalized to the peak value.
(b) Expanded view of the onset region for $h\nu = 5.2$--3.5 eV.
(c) Plots of onset kinetic energy versus photon energy from the $h\nu$-dependent HS-UPS spectra.}
\label{fig:LEHSUPS_Alq3}
\end{figure*}

Figure \ref{fig:LEHSUPS_Alq3}(a) shows the $h\nu$-dependent HS-UPS spectra of an Alq$_3$ thin film.
Photoelectron signals are observed for $h\nu > 2.1$ eV.
For $h\nu = 7.7$--6.5 eV, the spectral onset shifts to lower kinetic energy as the photon energy decreases (red bars).

At photon energies below the ionization energy ($h\nu < 6$ eV), two features are observed.
First, intense photoelectron signals appear near the secondary-electron cutoff, with nearly identical spectral lineshapes; their onsets are indicated by black bars in Figure \ref{fig:LEHSUPS_Alq3}(a).
Second, a very weak signal appears in the spectral onset region.
As shown in Figure \ref{fig:LEHSUPS_Alq3}(b), when this region is expanded, the onset of the weak signal shifts to lower kinetic energies with decreasing photon energy (blue bars in Figures \ref{fig:LEHSUPS_Alq3}(a) and (b)).
Figure \ref{fig:LEHSUPS_Alq3}(c) plots the onset kinetic energy as a function of photon energy.
For $h\nu = 7.7$--6.3 eV, the onset exhibits a linear dependence with a slope of approximately 1 (red plots).
Because this slope agrees with Eq. \ref{eq:SQEPEofHOMO}, these photoelectrons can be attributed to SQEPE from the HOMO.
For $h\nu < 5.4$ eV, the spectral onset (blue bars in Figures \ref{fig:LEHSUPS_Alq3}(a) and (b)) also exhibits a linear dependence with a slope of approximately 1 (blue plots in Figure \ref{fig:LEHSUPS_Alq3}(c)).
According to the model in the literature\cite{Kinjo2016_negativeion}, low-energy photon irradiation generates photo-carriers whose electrons are trapped near the surface due to positive polarization charges: 
\begin{align}
    \textrm{M}+h\nu\rightarrow \textrm{M}^*, 
    \label{eq:generate_photocarrier}\\
    \textrm{M}^* + \textrm{M} \rightarrow \textrm{M}^++\textrm{M}^-, 
    \label{eq:hot_electron_become_anion}
\end{align}
where $\textrm{M}^+$ and $\textrm{M}^-$ denote a cation and anion, respectively.
In the SQEPE process from the singly occupied molecular orbital (SOMO) of anion states, the photoelectron is excited according to 
\begin{equation}
E_\mathrm{k}^\mathrm{onset} = h\nu - \textit{I}_\mathrm{anion},
\label{eq:SQEPEofAnion}
\end{equation}
where $\textit{I}_\mathrm{anion}$ is the ionization energy from the anion state to the neutral state.
Because the observed linear relation with a slope near 1 (blue plots in Figure \ref{fig:LEHSUPS_Alq3}(c)) follows this equation, these photoelectrons are attributed to the SQEPE of anion states.
From the linear fit, the ionization energy of the anion states is estimated to be 2.40 eV.
This interpretation is further supported by the CFS-YS results \Add{(see Section \ref{sec:DOS})}.

Next, we focus on the intense photoelectron signals near the secondary-electron cutoff in Figure \ref{fig:LEHSUPS_Alq3}(a), indicated by black bars. Figure \ref{fig:LEHSUPS_Alq3}(c) shows that, remarkably, the kinetic energy of these onsets remains at 0.38 eV regardless of photon energy (black plots). This behavior does not follow the photon-energy dependence expected for SQEPE from the HOMO(Eq. \ref{eq:SQEPEofHOMO}), occupied in-gap states (Eq. \ref{eq:SQEPEofIngap}), or anion states , and (Eq. \ref{eq:SQEPEofAnion}).
To explain this spectral onset behavior, we consider BEE-related processes\cite{pope1999electronic_BEE_book, Ono2001_BEE}.
In the case of BEE via exciton--anion fusion, at first, an incident photon first generates a superexcited state ($\mathrm{M}^{**}$), which then rapidly relaxes to the first-excited state ($\mathrm{M}^{*}$) according to Kasha’s rule:
\[
\mathrm{M} + h\nu \rightarrow \mathrm{M}^{**} \rightarrow \mathrm{M}^{*}.
\]
Next, the exciton transfers its energy, $E(\mathrm{M}^\mathrm{*})$, to an anion, exciting an electron from the SOMO and ionizing the anion\Erase{ via a superexcited state}: 
\Add{
    \begin{equation}
        \mathrm{M}^{*} + \mathrm{M}^{-} \rightarrow 2\mathrm{M} + e^{-}.
        \label{eq:BEEofAnionSingletFusiuon}
    \end{equation}
    }
In this process, $E_\mathrm{k}^\mathrm{onset}$ is given by
\begin{align}
    E_\mathrm{k}^\mathrm{onset}=E(\mathrm{M}^\mathrm{*})-\textit{I}_\mathrm{anion},
    \label{eq:Ek_of_singlet_anion_fusion}
\end{align}
where $E(\mathrm{M}^{*})$ is the exciton energy. 
In the case of BEE via exciton--exciton fusion, the reaction proceeds as
\begin{equation}
    \mathrm{M}_1^* + \mathrm{M}_2^* \rightarrow \mathrm{M}_1 + \mathrm{M}_2^{**} \rightarrow \mathrm{M}_1 + \mathrm{M}_2^+ + e^{-},
\label{eq:BEE_exciton_exciton_fusion_reaction_formula}
\end{equation}
where $\mathrm{M}_1$ and $\mathrm{M}_2$ are two distinct molecules.
The onset kinetic energy of the resulting photoelectron is
\begin{align}
E_\mathrm{k}^\mathrm{onset} = E(\mathrm{M}_1^{*}) + E(\mathrm{M}_2^{*}) - \textit{I},
\label{eq:Ek_of_exciton_exciton_fusion}
\end{align}
where $E(\mathrm{M}_1^{*})$ and $E(\mathrm{M}_2^{*})$ are the excitation energies of the two excitons. 
For both BEE mechanisms---exciton--anion and exciton--exciton fusions---the spectral onset kinetic energy does not depend on the photon energy $h\nu$, consistent with the observed intense photoelectron signal at $E_\mathrm{k} < 0.38$ eV. 

To verify the contribution of BEE photoelectrons, we calculated $E_\mathrm{k}^\mathrm{onset}$ using Eqs. \ref{eq:Ek_of_singlet_anion_fusion} and \ref{eq:Ek_of_exciton_exciton_fusion}. 
The singlet and triplet energies were taken as the onsets of the photoluminescence spectrum (2.7 eV)\cite{Tanaka2005_Alq3_S1T1} and the phosphorescence spectrum (2.0 eV)\cite{Tanaka2005_Alq3_S1T1}, respectively. The calculated results are summarized in Table \ref{tab:exciton_fusion}. Based on these calculations, the observed photoelectrons at $E_\mathrm{k} < 0.38$ eV are attributed to BEE via singlet--anion fusion.
Other BEE processes are not observed because the photoelectron kinetic energy is insufficient to escape from the interior of the Alq$_3$ thin film.

\begin{table*}[htbp]
    \centering
    \begin{tabular}{lllc}
        \toprule
         Fusion& Reaction formula & Kinetic energy & Calculated value (eV)\\
        \midrule
        Singlet--singlet & $\mathrm{S}_1 + \mathrm{S}_1 \rightarrow \mathrm{M} + \mathrm{M}^{+} + e^-$ & $2E({\mathrm{S}_1}) - \mathit{I}_\mathrm{}$ &$-$0.62\\
        Triplet--triplet & $\mathrm{T}_1 + \mathrm{T}_1 \rightarrow \mathrm{M} + \mathrm{M}^{+} + e^-$ & $2E({\mathrm{T}_1}) - \mathit{I}_\mathrm{}$ & $-$2.02\\
        Singlet--triplet & $\mathrm{S}_1 + \mathrm{T}_1 \rightarrow \mathrm{M} + \mathrm{M}^{+} + e^-$ & $E({\mathrm{S}_1}) + E({\mathrm{T}_1}) - \mathit{I}_\mathrm{}$ &$-$1.32\\
        Anion--singlet & $\mathrm{M}^{-} + \mathrm{S}_1 \rightarrow 2\mathrm{M} + e^-$ & $E({\mathrm{S}_1}) - \mathit{I}_\mathrm{anion}$ &0.30\\
        Anion--triplet & $\mathrm{M}^{-} + \mathrm{T}_1 \rightarrow 2\mathrm{M} + e^-$ & $E({\mathrm{T}_1}) - \mathit{I}_\mathrm{anion}$ &$-$0.40\\
        \bottomrule
    \end{tabular}
    \caption{
    Possible biphotonic electron emission (BEE) processes via exciton--exciton and exciton--anion fusion, and the calculated maximum kinetic energies resulting from these processes. 
    $\mathit{I}$, 
    $\mathit{I}_\mathrm{anion}$, 
    $E(\mathrm{S}_1)$, and 
    $E(\mathrm{T}_1)$ denote 
    the ionization energy (6.02 eV) of a neutral molecule, 
    ionization energy (2.40 eV) of an anion, 
    singlet energy (2.7 eV)\cite{Tanaka2005_Alq3_S1T1}, and 
    triplet energy (2.0 eV)\cite{Tanaka2005_Alq3_S1T1}, respectively.
    }
    \label{tab:exciton_fusion}
\end{table*}

To verify our interpretation of the photoelectrons, we examined the relationship between the incident photon flux, $\textit{I}_\textrm{ph}$, and the photoelectron intensity.
At $h\nu = 3.06$ and 3.40 eV, the peak intensity scales approximately as $\textit{I}_\textrm{ph}^{1.7}$ and $\textit{I}_\textrm{ph}^{2.5}$, respectively
\Add{ (see Figure S1 in Supporting Information)}.
At these photon energies, photoelectrons are primarily generated via the BEE process through anion--singlet fusion, as discussed in Figure \ref{fig:LEHSUPS_Alq3}.
According to Eq. \ref{eq:BEEofAnionSingletFusiuon}, the photoelectron density, $[e_\textrm{ext}^-]$, is expressed as the product of the anion density, $[\textrm{M}^-]$, and the singlet exciton density, [$\mathrm{S}_1$]:
\begin{align} 
[\mathrm{e}^-_\mathrm{ext}] &\propto [\mathrm{M}^-][\mathrm{S}_1].\label{eq:density_of_BEE_anion_singlet}
\end{align}
Under light irradiation with $h\nu > E_\textrm{g}^\textrm{opt}$, the photon-flux dependence of the anion density is determined by the balance among the following anion-generating and anion-eliminating processes:
(i) An incident photon generates a photo-carrier, whose electron is subsequently captured near the surface by a molecule due to positive polarization charge, accumulating an anion (Eqs. \ref{eq:generate_photocarrier} and \ref{eq:hot_electron_become_anion}).
(ii) BEE via exciton--exciton fusion produces a hot electron (Eq. \ref{eq:BEE_exciton_exciton_fusion_reaction_formula}), which is then captured near the surface to accumulate an anion.
(iii) An anion is eliminated via photoionization through SQEPE.
(iv) An anion is eliminated via BEE through singlet--anion fusion (Eq. \ref{eq:BEEofAnionSingletFusiuon}).

Considering the balance of these processes, the anion density is expected, in principle, to scale between $\textit{I}_\textrm{ph}^0$ (constant) and $\textit{I}_\textrm{ph}^2$.
Meanwhile, the singlet density depends linearly on $\textit{I}_\textrm{ph}$ (Eq.~\ref{eq:density_of_BEE_anion_singlet}).
Consequently, the photoelectron density arising from BEE via anion--singlet fusion should scale between $\textit{I}_\textrm{ph}$ and $\textit{I}_\textrm{ph}^3$.
In practice, excitons have a finite lifetime, and some decay independently of BEE processes.
As a result, the photoelectron intensity follows $\textit{I}_\textrm{ph}^{\alpha}$ with $\alpha < 3$, consistent with the observed values.

In contrast, under light irradiation at $h\nu = 2.36$ eV, the photoelectron intensity increases linearly with $\textit{I}_\textrm{ph}$ \Add{(see Figure S1 in Supporting Information)}. \Add{Because this photon energy is below the optical bandgap, excitons are not produced. Consequently, the photoelectrons due to BEE effect are not generated, whereas only photoelectrons due to SQEPE from anions are observed.} These photon-flux dependencies at $h\nu = 3.40$, 3.06, and 2.36 eV support our assignment of the observed photoelectrons to SQEPE of anions and BEE effect via anion--singlet fusion. \Add{Furthermore, the order-of-magnitude estimate suggests that the anion and exciton populations are a realistic amount to observe the BEE signal. The detail is described in Support Information.}
\subsection{Density of states}\label{sec:DOS}

\begin{figure}
    \includegraphics[width=85 mm]{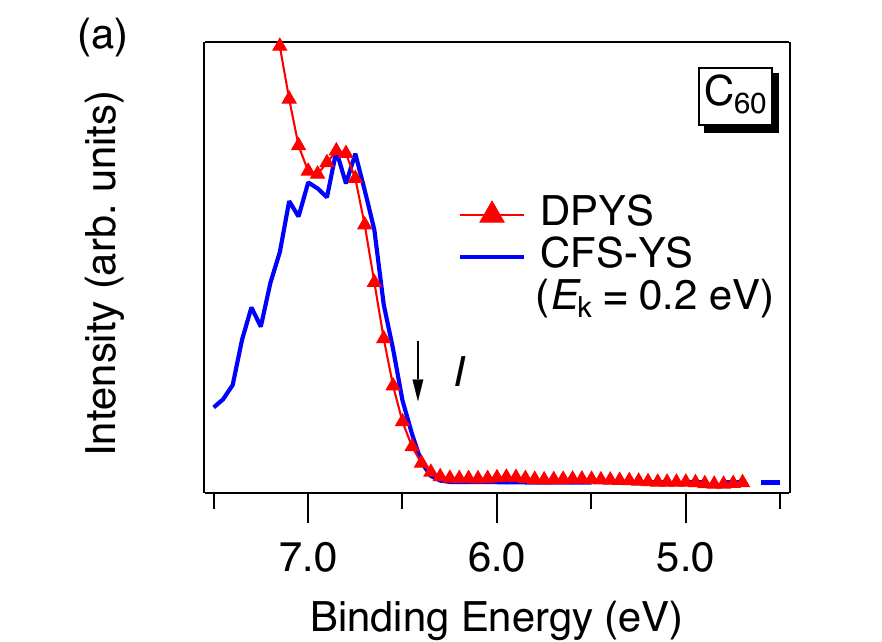}
    \includegraphics[width=88 mm]{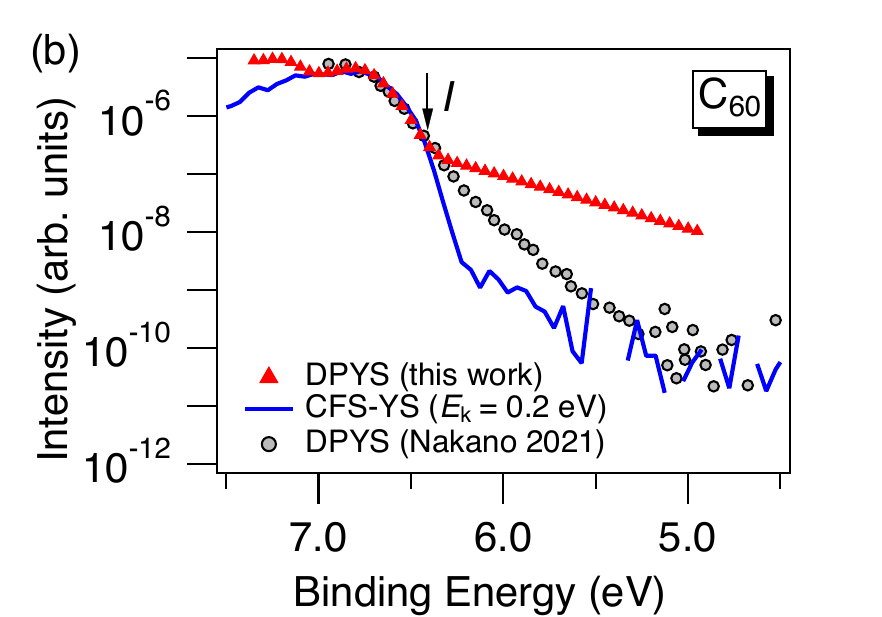}
    \caption{
    Comparison of the DOS determined by DPYS and CFS-YS for C$_{60}$ shown in (a) linear and (b) logarithmic scales. CFS-YS was performed at the peak of the secondary-electron cutoff ($E_\mathrm{k} = 0.20$ eV). \Add{DPYS obtained by Nakano \textit{et al}. is also plotted.\cite{Nakano2021_PYS_differentiation}}
    }
    \label{fig:DPYS_CFS_C60}
\end{figure}

As discussed above, photoelectrons in organic materials can originate not only from SQEPE of the HOMO and in-gap states, but also from SQEPE of anion states and the BEE effect. In this section, we establish a method to reliably determine the DOS of in-gap states by using PYS and CFS-YS.

We first analyzed the DOS of in-gap states in C$_\mathrm{60}$, where SQEPE from the HOMO and in-gap states is observed.
CFS-YS was performed at $E_\mathrm{k} = 0.20$ eV, following the common practice of measuring at the secondary-electron cutoff peak to maximize the signal-to-noise ratio. 
Figures \ref{fig:DPYS_CFS_C60}(a) and (b) show the CFS-YS spectra in linear and logarithmic scales, respectively, with binding energy on the horizontal axis and the vacuum level set to 0 eV. 
The DOS of the HOMO appears at $h\nu > 6.3$ eV. On a logarithmic scale, the DOS of HOMO follows a parabolic curve, indicating a Gaussian distribution. In-gap states are observed between $h\nu = 6.2$--5.5 eV. 

\Add{Next, the PYS spectrum was differentiated to estimate the DOS distribution. 
Because the signal-to-noise ratio of the PYS spectrum in the bandgap region was insufficient for differentiation, the PYS spectrum below 6.4 eV was first fitted with an exponential function, and the fitted curve was used in place of the experimental data. 
Then, the spectrum was smoothed by convolution with a Gaussian function with a full width at half maximum of 0.2 eV and subsequently differentiated.}
    
\Add{On both linear and logarithmic scales, the HOMO DOS lineshape obtained from DPYS agrees well with that from CFS-YS. 
In contrast, the DOS of in-gap states differs between the two methods. 
DPYS shows in-gap states in the range of 6.2–5.0 eV with an exponential tail, \ReviseII{whose density is higher and slope is larger than those obtained by CFS-YS}{however, its density and lineshape differ from that obtained by CFS-YS}.
We additionally compared the DPYS lineshape reported in the literature\cite{Nakano2021_PYS_differentiation} with the CFS-YS spectrum. 
The reported DPYS spectrum also exhibits \ReviseII{a higher density in the bandgap region than that of CFS-YS}{ in-gap states, but its density and lineshape are also different from those of CFS-YS}.}

\EraseII{From these findings, we conclude that DPYS can provide a rough estimation of the DOS distribution of in-gap states, whereas CFS-YS obtains a more reliable DOS lineshape. This difference likely arises because CFS-YS can determine the DOS with fewer assumptions than PYS (see Section~\ref{sec:theory})}\AddII{This difference likely arises from the assumptions required by DPYS. In the DPYS technique, low-energy photoelectrons must be analyzed to obtain the DOS profile of in-gap states. Furthermore, DPYS requires the assumption that the density of final states ($D_\textrm{f}$), transport probability ($X$), and transmittance probability ($T$) are constant (see Eqs. \ref{eq:PYS} and \ref{eq:PYS_int_DidE}) to regard its lineshape as density of initail states ($D_\textrm{i}$), however, basically these terms depend on photoelectron kinetic energy in the low-energy region. On the other hand, in the CFS-YS method, these terms are constant since this method detects photoelectrons at a fixed kinetic energy (see Eq. \ref{eq:CFSYS}). From these findings, we conclude that DPYS can provide a rough estimation of the DOS distribution of in-gap states, whereas CFS-YS can obtain a DOS profile with fewer methodological assumptions.}

\begin{figure}
    \includegraphics[width=75mm]{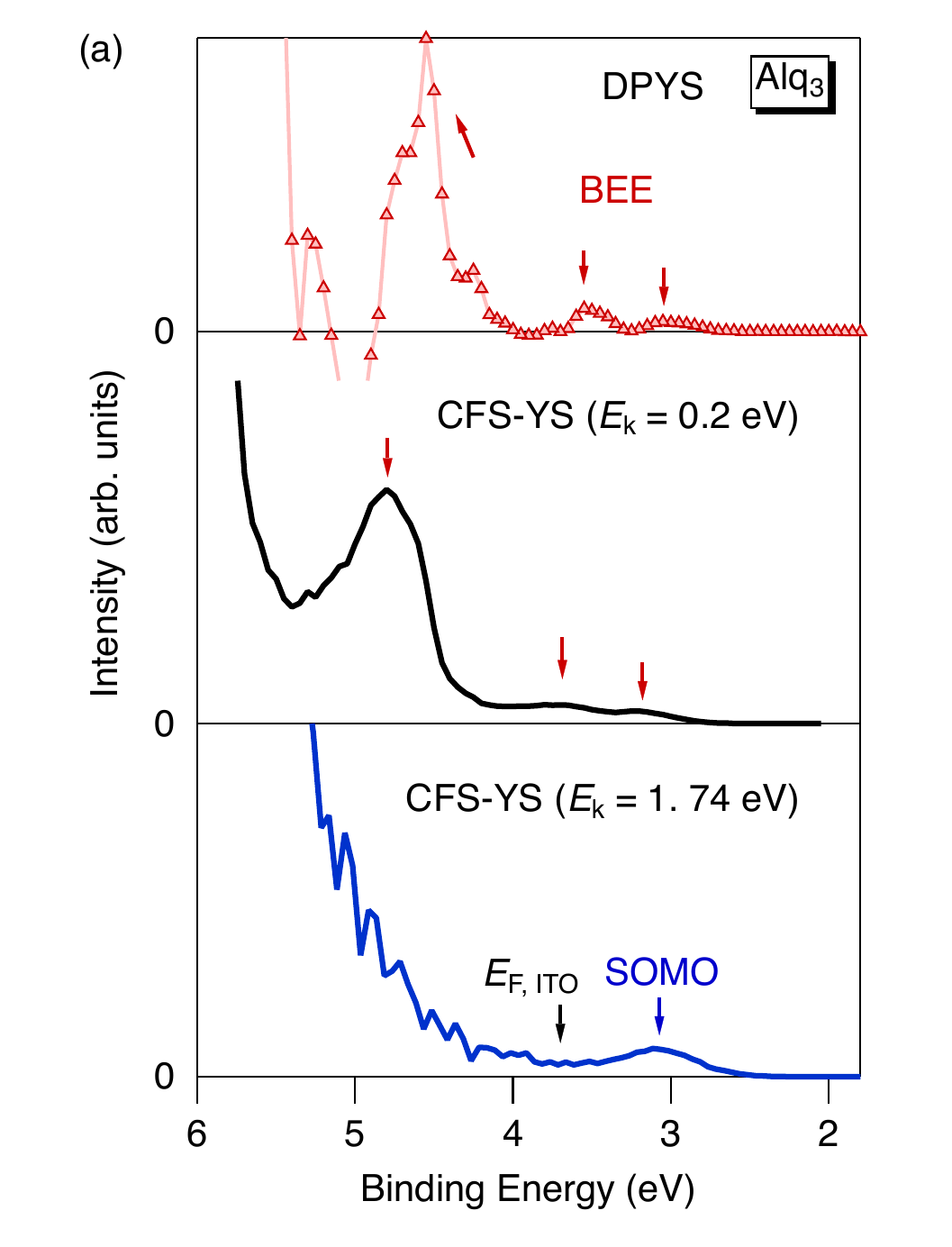}
    \includegraphics[width=75mm]{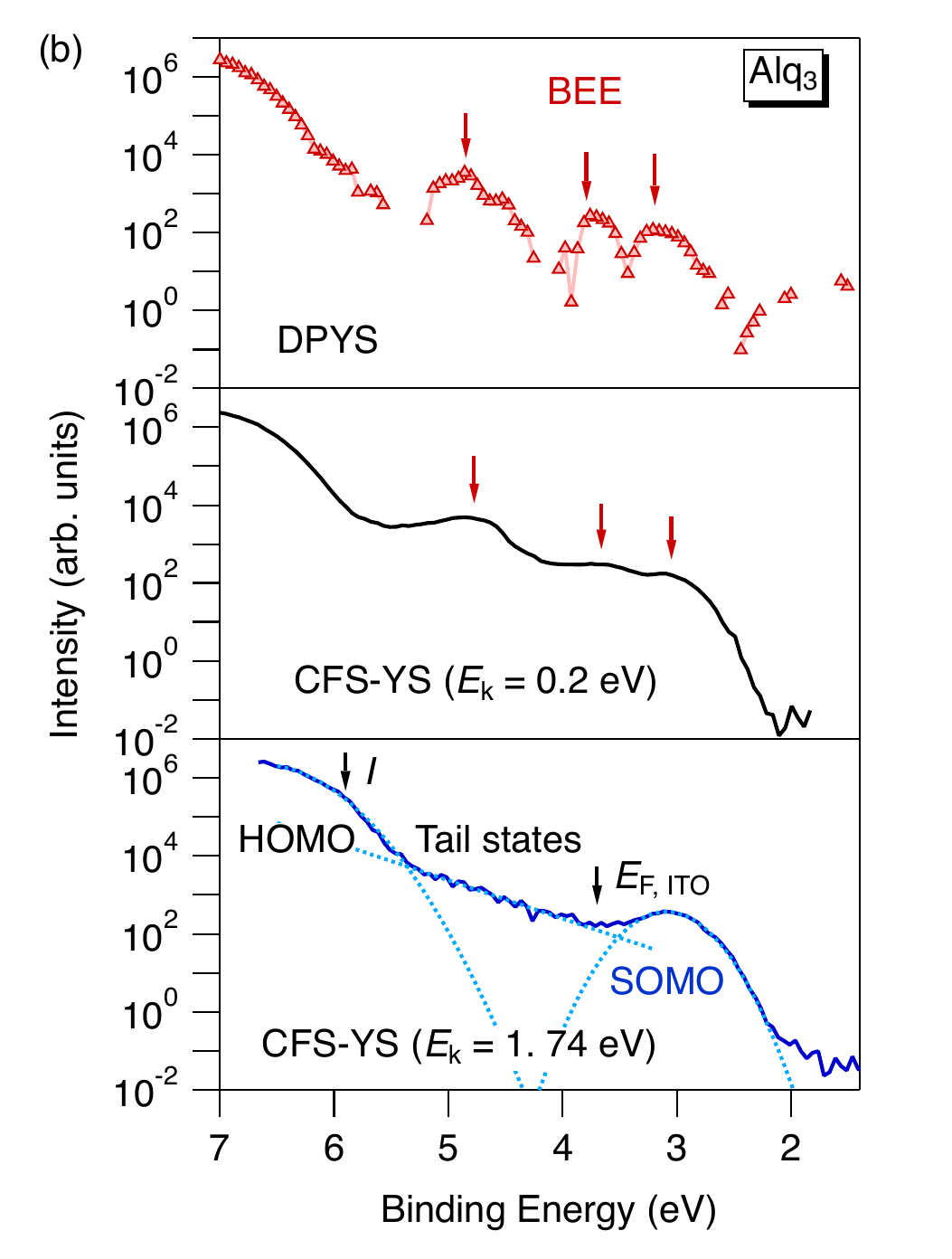}
    \caption{
    Comparison of the DOS determined by DPYS and CFS-YS for Alq$_{3}$ in (a) linear and (b) logarithmic scales. CFS-YS was performed at different kinetic energies, $E_\mathrm{k} = 0.20$ and 1.74 eV. Peaks arising from biphotonic electron emission (BEE) are indicated by red arrows.
    \Add{$I$ and $E_\textrm{F, ITO}$ represent ionization energy and Fermi level of ITO substrate, respectively.}
    }
    \label{fig:DPYS_CFS_Alq3}
\end{figure}

Figure \ref{fig:DPYS_CFS_Alq3}(a) and (b) show the DPYS spectrum of an Alq$_3$ thin film in linear and logarithmic scales, respectively.
DPYS exhibits oscillations around 4.6, 3.5, and 3.1 eV and shows negative values at $h\nu = 5.5$--5.2 eV and 3.8--4 eV, indicating that DPYS does not accurately estimate the DOS.

Next, we performed CFS-YS using the traditional procedure, in which the partial photoelectron yield is measured at the secondary-electron cutoff peak to maximize the signal-to-noise ratio.
Photoelectrons at $E_\mathrm{k} = 0.20$ eV were detected, and the resulting CFS-YS spectrum exhibits peaks at approximately 4.5, 3.5, and 3.0 eV, similar to the DPYS spectrum.
Because photoelectrons with $E_\mathrm{k} < 0.38$ eV are primarily dominated by BEE \Add{(see Section \ref{sec:ups})}, the CFS-YS spectrum below the ionization energy does not reflect the DOS of in-gap states but rather the BEE signal under the conventional measurement procedure.
These results indicate that conventional CFS-YS and PYS cannot reliably determine the DOS distribution of in-gap states in organic semiconductors where BEE contributes to the photoelectron signal. 

To reliably determine the DOS of in-gap states, CFS-YS was performed at a higher kinetic energy, $E_\mathrm{k} = 1.74$ eV, well above the BEE-dominated region ($E_\mathrm{k} < 0.38$ eV). In this article, this measurement is referred to as “high-$E_\mathrm{k}$ CFS-YS” to distinguish it from the conventional CFS-YS.
In high-$E_\mathrm{k}$ CFS-YS, the spectral peaks arising from BEE disappear, allowing the DOS lineshape of in-gap states to be accurately determined.
The DOS of in-gap states follows an exponential distribution, $D(E_\mathrm{b})\propto \exp(-(E_\mathrm{b}-E_\textrm{v})/E_0)$, where $E_\mathrm{b}$ and $E_\mathrm{v}$ are the binding energy and valence-band maximum, respectively, and the slope parameter $E_0$ is 0.3 eV.

In addition to the exponential in-gap states, the CFS-YS spectrum shows a peak at $E_\mathrm{b} = 3.1$ eV, which follows a Gaussian profile.
This feature is assigned to the DOS of the SOMO of an anion (electrons occupying the LUMO of the neutral molecule). 
The DOS lineshape of the SOMO is revealed here by the high-$E_\mathrm{k}$ CFS-YS measurement in contrast to the previous study\cite{Kinjo2016_negativeion}. 
The onset energy of the DOS of SOMO is 2.5 eV. 

\Add{The effective work function (the difference between the Fermi level of the ITO substrate and the vacuum level of Alq$_3$) is 3.7 eV, which corresponds to the Fermi level of the ITO in the binding energy axis in Figure \ref{fig:DPYS_CFS_Alq3}.
SOMO states are observed in the region shallower than the Fermi level of ITO, indicating that the Fermi levels of Alq$_3$ and ITO do not correspond. This observation of photoelectron shallower than the Fermi level is also often observed in other organic materials\cite{Sato2017_HSUPS_1015cm-3eV-1}. The details of this mechanism are beyond the scope of this study.}

The SOMO peak in the high-$E_\mathrm{k}$ CFS-YS spectrum appears at 3.1 eV, whereas the first peak reflected LUMO in low-energy inverse photoelectron spectroscopy (LEIPS) is reported at around 1.5 eV\cite{Yoshida2015_OLED_LEIPS}.
This shift indicates that electrons in the SOMO are more stabilized relative to those in the LUMO. 
The discrepancy arises from differences in the detection mechanisms of the two methods.
LEIPS probes the LUMO via a transition from the neutral state to an anion, with the observed LUMO primarily influenced by electronic polarization.
In contrast, high-$E_\mathrm{k}$ CFS-YS probes the SOMO through a transition from an anion to a neutral state, which is identical to anion photoelectron spectroscopy. Since the anions near the surface are in a steady state, the observed SOMO reflects not only electronic polarization but also slower relaxation processes, such as ionic and oriented polarization effects.

The energetic difference between the LUMO  and SOMO corresponds to the reorganization energy, $\lambda = \mathrm{VDE}-\mathrm{VEA}$, where VDE and VEA are the vertical detachment energy and vertical electron affinity, respectively. 
Although precise determination of VDE and VEA is challenging, taking the SOMO peak of the high-$E_\mathrm{k}$ CFS-YS spectrum as VDE and the first peak of LEIPS as VEA gives an estimated surface reorganization energy, $\lambda_\text{surface}$ $\sim$1.6 eV.
This value is larger than the bulk reorganization energy of approximately 0.7 eV deduced by anion photoelectron spectroscopy for gas phase Alq$_3$ in the literature\cite{Yanase2018_reorganization_energy_alq3}. 
The difference arises because anion photoelectron spectroscopy probes gas-phase anions, whereas high-$E_\mathrm{k}$ CFS-YS probes anions near the surface of thin films, where spontaneous orientation polarization occurs\cite{Kinjo2016_negativeion}.
In the thin film, anions are accumulated by positive polarization charges near the surface.
Consequently, anions at the surface are energetically more stabilized than those in the bulk due to Coulomb interactions between the SOMO electron and the positive polarization charges. 

In OLED devices, Alq$_3$ is commonly used as the emission layer or electron transport layer.
At the interfaces between Alq$_3$ and a cathode or electron injection layer, Alq$_3$ molecules form anion states due to electron trapping by positive polarization charges\cite{Tanaka2019_interface_alq3_cath, Noguchi2013_interface_alq3_cath}.
Notably, the observation that the surface SOMO is more stabilized than the LUMO aligns qualitatively with previous studies\cite{Tanaka2019_interface_alq3_cath, Noguchi2013_interface_alq3_cath, Altazin2016_simulation_of_OLED}, which reported that positive polarization charges in Alq$_3$ reduce the electron injection barrier from the electrode, thereby enhancing injection efficiency. 
Thus, evaluating only the LUMO of neutral Alq$_3$ via LEIPS is insufficient to fully understand carrier behavior in devices.
It is essential to combine information on the LUMO and SOMO, determined by LEIPS and high-$E_\mathrm{k}$ CFS-YS, respectively.

\begin{figure}
\centering
\includegraphics[width=83mm]{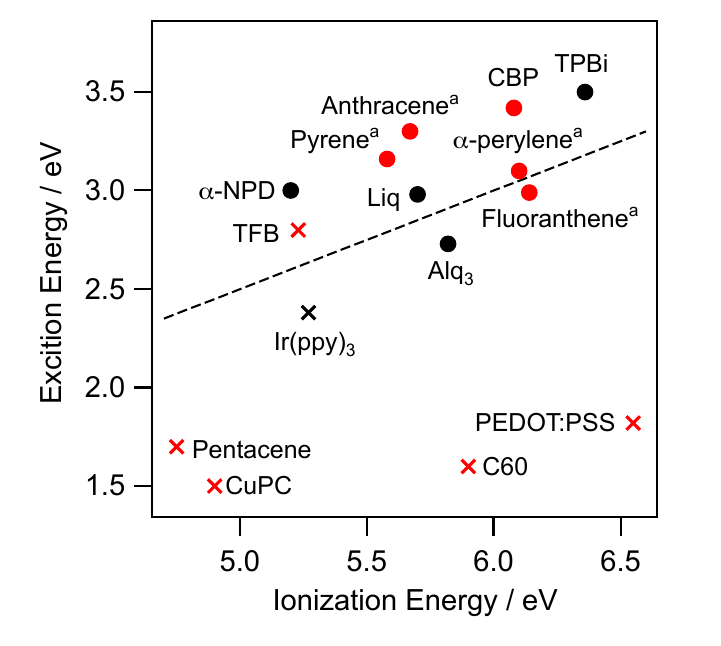}
    \caption{
            Plot of ionization energy and exciton energy in organic semiconductors.
            Exciton energy is approximated by the optical bandgap. The dashed line indicates the expected subthreshold for observing photoelectrons from BEE via singlet--singlet fusion ($2E(\textrm{S}_1) - I = 0$). 
            Circles represent materials where BEE-induced photoelectrons were observed, and crosses represent materials where they were not observed. \ReviseII{Red and black symbols represent nonpolar and polar molecules, respectively.}
            {Red symbols represent nonpolar molecules and polymers, whereas black symbols represent polar molecules.}
            Data marked with $^{a}$ are from the literature\cite{Ono2001_BEE}.}
  \label{fig:BEE_Ith_Egopt}
\end{figure}

\Erase{Photoelectrons arising from BEE are observed not only in Alq$_3$ but also in many other organic materials. }\Add{To examine whether BEE is generally observed, we widely investigated a series of organic semiconductors.} Figure \ref{fig:BEE_Ith_Egopt} shows the relationship between exciton energy and ionization energy.
Exciton energy was approximated by the optical bandgap as a first-order estimate of the singlet energy.
Ionization energy was determined using $h\nu$-dependent HS-UPS or PYS.
Circles indicate materials in which BEE was observed, while crosses indicate materials in which it was not. 
\ReviseII{Red and black symbols represent nonpolar and polar molecules, respectively.}
{Red symbols represent nonpolar molecules and polymers, whereas black symbols represent polar molecules.}
The dashed line represents the threshold for BEE via singlet--singlet fusion: $2E(\mathrm{S}_1)-I = 0$ eV.
BEE is expected to occur when $2E(\mathrm{S}_1)-I>0$ and not to occur when $2E(\mathrm{S}_1)-I<0$. 

In most materials, the observed results are consistent with this simple classification, supporting that BEE via singlet--singlet fusion is widely observed in organic semiconductors. 
Alq$_3$ lies below the dashed line in the plot, yet BEE via anion--singlet fusion is observed. Fluoranthene also lies below the line, but a previous study reported BEE via singlet--singlet fusion\cite{Ono2001_BEE}. \Add{This article states that} this discrepancy may arise from experimental errors in the reported ionization energy.
Overall, these results indicate that BEE effect is a common phenomenon in organic semiconductors. 
\Add{Furthermore, we found that BEE is observed regardless of the polarity of the molecules. Because nonpolar molecules do not exhibit surface orientation polarization, surface orientation polarization is unrelated to the observation of BEE-induced photoelectrons.}
    
Based on these findings, we establish guidelines for reliably determining the DOS of in-gap states.
In PYS and CFS-YS measurements, low-kinetic-energy photoelectrons due to SQEPE are analyzed to probe the DOS of in-gap states. 
However, measurement photons frequently also produce intense low-kinetic-energy photoelectrons due to BEE. 
Such BEE signals obscure contributions from SQEPE of the in-gap states. 
\Add{In this case, because PYS measures the total photoelectron yield without selecting photoemission kinetic energy, the BEE contribution cannot be eliminated from the PYS spectrum, causing DPYS spectra to misestimate the true DOS.
In contrast to PYS technique, CFS-YS measures the partial yield at the specific kinetic energy. However, conventional CFS-YS also does not reflect the DOS of in-gap states because it measures partial yield near secondary-electron cutoff, where BEE contributions are dominated. 
By measuring partial yields at kinetic energies higher than that of BEE signals, CFS-YS can provide a reliable DOS profile of the in-gap states without BEE contributions.
In Alq$_3$, this approach additionally reveals the DOS of the SOMO states.}

\Add{To ensure that the DPYS or CFS-YS spectra reflect the DOS of in-gap states, it is effective to examine the photon flux dependence of the spectral intensity at several photon energies. 
If photoelectron intensity reflects occupied DOS, its intensity depends on the incident photon flux. In contrast, if photoelectron intensity reflects BEE, its intensity is not proportional to incident photon flux (\textit{e.g.}, BEE due to the singlet–anion fusion in Alq$_3$ in Figures S1(a) and (b)). 
In exciton–exciton fusions, in principle, BEE signals are expected to be proportional to the square of the photon flux, as indicated by the reaction formula in Table \ref{tab:exciton_fusion}.}

\Add{In addition to photon-flux dependence,} at a minimum, the expected kinetic energy of BEE should be calculated using the equations summarized in Table \ref{tab:exciton_fusion}. If all BEE processes yield negative kinetic energies, BEE should not be observed in principle.
If BEE is present, its contribution can be identified in the measured spectral lineshape: the BEE threshold corresponds to the singlet or triplet excitation energies, so changes in the spectral lineshape or the spectral onset are expected at these energies. 

\Erase{
    In contrast to DPYS and conventional CFS-YS, measurements cannot reflect the DOS of HOMO and in-gap states, but the BEE signal. 
    High-$E_\mathrm{k}$ CFS-YS provides a reliable determination of the DOS of the HOMO and occupied in-gap states by avoiding contributions from the BEE signal. 
    In Alq$_3$, this approach also reveals the DOS of the SOMO.
    In contrast to the CFS-YS approach, in principle, PYS measurement cannot eliminate BEE contribution because this method cannot resolve photoelectron yield by energy axis.
    }

These findings establish practical guidelines for measuring in-gap states using low-energy photons 
as follows:
\Add{(i) To check whether a measured spectrum originates from the occupied DOS or from the BEE, measure the photon-flux dependence of the photoelectron intensity at several photon energies}
(ii) and calculate the expected kinetic energy of photoelectrons resulting from BEE using Table \ref{tab:exciton_fusion}. 
(iii) If the spectrum is likely dominated by BEE, employ high-$E_\mathrm{k}$ CFS-YS to obtain a reliable DOS.

\subsection{Effect of BEE on optoelectronic devices}
\begin{figure}
  \centering
  \includegraphics[width=85mm]{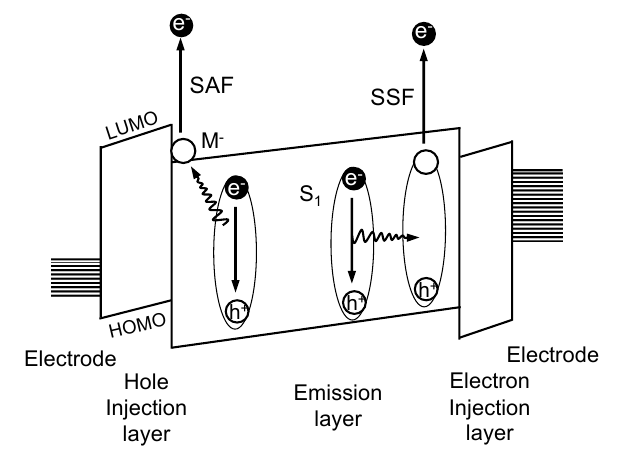}
  \caption{Effect of BEE via exciton--exciton fusion on OLED device performance. SSF and SAF denote singlet--singlet fusion and singlet--anion fusion, respectively.}
  \label{fig:BEEinDevice}
\end{figure}

Finally, we discuss the impact of BEE on OLED performance using Figure \ref{fig:BEEinDevice}. 
The BEE is caused by exciton--exciton and exciton--anion fusion processes. 
In the luminescent layer, where many excitons exist, exciton--exciton fusion (Eq. \ref{eq:BEE_exciton_exciton_fusion_reaction_formula}) is likely to occur. 

\Erase{
    When a singlet relaxes to a neutral molecule in exciton--exciton fusion processes by transferring its singlet energy to the other exciton, this fusion can act as a non-radiative deactivation process, reducing luminescence efficiency. 
    For example, in singlet–singlet fusion processes, a singlet relaxes to a neutral molecule by transferring its singlet energy to the other singlet, which can be interpreted as a non-radiative deactivation process, reducing luminescence efficiency. 
    The other exciton separates a cation and a hot electron via a superexcited state by receiving the singlet energy.
    This hot electron may induce bond dissociation. Additionally, the generated cation can function as a degradation species.
    }
\Add{For example, in singlet–singlet fusion process, a singlet gives its energy to the other singlet, which is interpreted as a non-radiative deactivation process. 
The other exciton, which accepts that energy, can be photoionized via a superexcited state, leading to charge separation into a cation and a hot electron (Eq. \ref{eq:BEE_exciton_exciton_fusion_reaction_formula}). 
In blue thermally activated delayed fluorescence OLEDs, singlet–singlet fusion is likely the dominant degradation pathway due to the photoionization-induced decomposition\cite{Liu2018_BEEdevice}. 
In blue phosphorescent OLEDs, singlet–singlet fusion is proposed as an origin of device degradation due to molecule dissociation\cite{Scholz2015_BEEdevice)}. 
}\Erase{The high-energy hot electron, having energy well above the LUMO of the hole injection layer, can be easily injected into it under an applied bias, causing increased power consumption and a disruption of carrier balance.}

In OLEDs to which a bias is applied, electrons can accumulate at the heterointerfaces of the emission layer, forming anions. 
Consequently, singlet--anion fusion (Eq. \ref{eq:BEEofAnionSingletFusiuon}) occurs in the luminescence layer. Singlet--anion fusion can also be a non-radiative deactivation process because the singlet deactivates with the transfer of its singlet energy to the anion. This fusion can be the origin of the decrease in photoluminescence efficiency above the turn-on voltage in OLEDs due to the electron accumulation at the interface\cite{Noguchi2022_excition_quench}.\Erase{ Moreover, the BEE process can also be viewed as a carrier-generation mechanism in organic solar cells, as exciton fusion produces both a hot electron and a cation (Eq. \ref{eq:BEE_exciton_exciton_fusion_reaction_formula}). This carrier generation may influence solar-cell performance.} The $h\nu$-dependent HS-UPS technique provides direct access to the kinetic energy distribution of photoelectrons arising from BEE via exciton fusions, offering a powerful tool for detailed investigation of exciton dynamics in OLEDs. 

\section{Summary}
This study investigated the origin of photoelectrons in organic semiconductors using PYS, $h\nu$-dependent HS-UPS, and CFS-YS to establish a reliable method for determining the DOS of in-gap states. 
We found that low-energy photon irradiation widely induces BEE in organic semiconductors. Under these conditions, derivative PYS fails to accurately determine the DOS due to the intense BEE signal.
Similarly, conventional CFS-YS, which measures the partial photoelectron yield at the secondary-electron cutoff, also cannot reliably determine the DOS because of BEE contributions. 
\Add{In both techniques, it is effective to check the photon-flux dependence of spectral intensity at several photon energies to distinguish whether the observed signals arise from the DOS or BEE.}

In contrast, we demonstrated that performing CFS-YS at a kinetic energy higher than that of BEE allows the reliable determination of the DOS, including the in-gap states and, in the case of Alq$_3$, the SOMO of anions. 
For Alq$_3$ thin films, high-$E_\mathrm{k}$ CFS-YS measurements revealed the DOS of the HOMO, in-gap states, and the SOMO. 
The SOMO peak and onset energies were determined to be 3.1 eV and 2.6 eV, respectively, which are more stabilized than the LUMO observed by LEIPS in previous studies. 
This stabilization indicates that positive polarization charges arising from spontaneous orientation polarization facilitate electron injection from electrodes into Alq$_3$. 
Furthermore, the BEE process should be considered a potential carrier-generation and degradation pathway in optoelectronic devices. 
Both $h\nu$-dependent HS-UPS and high-$E_\mathrm{k}$ CFS-YS are demonstrated to be powerful techniques for probing exciton--exciton and exciton--anion interactions, which are key contributors to device performance, by selectively detecting the BEE signal.


\begin{acknowledgement}

This work was supported by JSPS KAKENHI Grant Number 25KJ0391.



\end{acknowledgement}


\bibliography{ref}

\end{CJK*}
\end{document}